\documentstyle[preprint,floats,pra,aps,epsfig,tighten]{revtex}
\begin{document} \draft

\title{\bf  Feynman's Decoherence}

\author{Y. S. Kim\footnote{electronic address: yskim@physics.umd.edu}}
\address{Department of Physics, University of Maryland,
College Park, Maryland 20742, U.S.A.}

\author{Marilyn E. Noz\footnote{electronic address: noz@nucmed.med.nyu.edu}}
\address{Department of Radiology, New York University,
New York, New York 10016, U.S.A.}

\maketitle

\begin{abstract}
Gell-Mann's quarks are coherent particles confined within a hadron at
rest, but Feynman's partons are incoherent particles which constitute
a hadron moving with a velocity close to that of light.  It is
widely believed that the quark model and the parton model are two
different manifestations of the same covariant entity.  If this is the
case, the question arises whether the Lorentz boost
destroys coherence.  It is pointed out that this is not the case, and
it is possible to resolve this puzzle without inventing new physics.
It is shown that this decoherence is due to the measurement processes
which are less than complete.
\end{abstract}

\narrowtext

\section{Introduction}\label{intro}
The superposition principle is the backbone of quantum mechanics.
We are thus led to think the transition from quantum to classical
mechanics is accompanied by a loss of coherence.  The transition from
classical to quantum mechanics thus requires that particles somehow
gain coherence.  Indeed, these have been outstanding problems
ever since the present form of quantum mechanics was formulated in
the late 1920s.

We are quite familiar with coherent and incoherent light waves.
If the phases become random, we say that the light waves are
incoherent.  Thus, we are naturally led to say that the process of
randomizing phases is a form of decoherence.  During this process, the
entropy of the system increases.  We know also that the transition
from quantum to classical mechanics has a deeper content than
randomizing phases.   Indeed, we still have to define the precise
meaning of the word ``decoherence.''  In the meantime,
we have to review critically the existing processes where quantum
coherence disappears.

In this report, we examine a concrete case of Feynman's parton picture.
The question is very simple.  According to Gell-Man~\cite{gell64},
hadrons like protons and mesons, are bound states of quarks capable of
quantum excitations like the hydrogen atom.  This picture is valid if the
hadrons are in their rest frame or move slowly.  If the hadron moves
very fast, it appears as a collection of free independent particles.
This observation was made first by Feynman, who named those free
particles as ``partons''~\cite{fey69}.

It is a widely accepted view that the quark model and the parton model
are two different manifestations of one covariant entity, with the
quark model valid when the hadron is at rest and the parton model
observed for the hadron moving with a speed close to that of light.
Quarks are coherent and partons are incoherent.  Then does the Lorentz
boost wash out the coherence as in randomizing phase differences for
light waves?  No!  Because we are not adding any physical processes.
We are simply looking at the same thing from different Lorentz frames.
Therefore, we are led to conclude that there are limitations on our
observation processes.

The parton picture was invented by Feynman based on phenomenology in
high-energy physics.  In order to appreciate fully its implications,
it is important to study his papers written on other subjects.  We also
should take into account the fact that Feynman could come up with
creative physical ideas without mathematical rigor or sometimes with
wrong mathematics.  Indeed, in order to understand the parton picture,
we need a broader physical and mathematical base.

The issue is that Feynman's original parton picture is valid only for
fast-moving hadrons.  Is it possible to construct a covariant model
which produces the parton model as a limit?  We can ask the same question
about Gell-man's quark model.  It is based on non-relativistic quantum
mechanics of bound states.  Is there a covariant picture of the quantum
bound state?  Indeed, Feynman raised this question and published a paper
on this subject with Kislinger and Ravndal~\cite{fkr71}.

Feynman was also interested in the concept of entropy coming from
measurement processes which are less than complete~\cite{fey72}.  This
subject was of course originated by von Neumann in his classic book
on mathematical foundations of quantum mechanics~\cite{neu32}.  There
are many different forms of measurement theory, and they are currently
debated in the literature in connection with the EPR problems and
Bell's inequality.  The letter "E" in EPR stands for Einstein.  Then
we should take into account another contribution made by Einstein,
namely Lorentz covariance.  It is therefore essential to study the
entropy effects when we deal with the covariance, because our
the measurement theory for non-relativistic quantum mechanics is
necessarily incomplete if we introduce another variable, namely the
time variable.

We shall consider all of the above items in this report.  We shall
start with Feynman's research style in Sec.~\ref{feynm}.  Feynman
was quite fond of using harmonic oscillators to find a clue to
new physics.  In Sec.~\ref{coupled}, we use coupled harmonic
oscillators to illustrate Feynman's physical ideas.  The oscillators
are useful in illustrating Feynman's rest of the universe and Feynman's
relativistic oscillators.

In Sec.~\ref{restof}, it is noted that
the time-separation variable does not exist in the present form of
quantum mechanics, while it clearly exists physically wheresoever the
space-like separation, like the Bohr radius, exists if we believe in
Einstein's special relativity.
If the system is Lorentz-boosted, the time-separation variable becomes
prominent because it is linearly mixed with the space-separation variable.
If we do not measure this variable, there will be an increase in entropy.

In Sec.~\ref{par}, we consider a hadron consisting of two quarks bound
together by a harmonic oscillator potential.  It is shown that
the covariant harmonic oscillator formalism gives non-relativistic
quantum mechanics in the Lorentz system where the hadron is at rest.  It
is then shown that the same formalism produces Feynman's parton model
in the frame where the hadron moves with a speed close to that of light.
Finally, in Sec.~\ref{measure}, we study carefully why the partons
appear as incoherent particles.  It is shown that it is due to our
measurement process which is less than complete in both the rest
frame and in the frame in which the hadron moves with a speed close
to that of light.

\section{Feynman's World}\label{feynm}
Feynman was quite fond of using harmonic oscillators to probe new
territories of physics.  In this section, we examine which
oscillator formalism is most suitable to interpret some of
Feynman's papers during the period 1969 -- 1972.  This formalism
should accommodate special relativity and quantum mechanics of
extended objects.

Let us start with a simple physical system.
Two coupled harmonic oscillators play many important roles in physics.
In group theory, it generates symmetry group as rich as
$O(3,3)$~\cite{hkn95jm}.  It has many interesting subgroups useful
in all branches of physics.  The group $O(3,1)$ is of course essential
for studying covariance in special relativity.  It is applicable to
three space-like variables and one time-like variable.  In the
harmonic oscillator regime, those
three space-like coordinates are separable.  Thus, it is possible to
separate longitudinal and transverse coordinates.  If we leave out
the transverse coordinates which do not participate in Lorentz boosts,
the only relevant variables are longitudinal and time-like variables.
The symmetry group for this case is $O(1,1)$ easily derivable from
the Hamiltonian for the two-oscillator system.

It is widely known that this simple mathematical device is the basic
language for two-photon coherent states known as squeezed states of
light~\cite{yuen76,knp91}.
However, this $O(1,1)$ device plays a much more powerful role in physics.
According to Feynman, {\it the adventure of our science of physics is a
perpetual attempt to recognize that the different aspects of nature
are really different aspects of the same thing}~\cite{feyaip}.  Feynman
wrote many papers on different subjects of physics, but they are
coming from one paper according to him.  We are not able to combine
all of his papers, but we can consider three of his papers published
during the period 1969-72.

In this paper, we would like to consider Feynman's 1969 report on
partons~\cite{fey69}, the 1971 paper he published with his students
on the quark model based on harmonic oscillators~\cite{fkr71}, and
the chapter on density matrix in his
1972 book on statistical mechanics~\cite{fey72}.  In these three
different papers, Feynman deals with three distinct aspects of nature.
We shall see whether Feynman was saying the same thing in these
papers. For this purpose, we shall use the $O(1,1)$ symmetry derivable
directly from the Hamiltonian for two coupled
oscillators~\cite{hkn99ajp}.  The standard procedure for this
two-oscillator system is to separate the Hamiltonian using normal
coordinates.  The transformation to the normal coordinate system
becomes very simple if the two oscillators are identical.  We shall
use this simple mathematics to find a common ground for the
above-mentioned articles written by Feynman.

First, let us look at Feynman's book on statistical
mechanics~\cite{fey72}.  He makes the following statement about the
density matrix. {\it When we solve a quantum-mechanical problem, what we
really do is divide the universe into two parts - the system in which we
are interested and the rest of the universe.  We then usually act as if
the system in which we are interested comprised the entire universe.
To motivate the use of density matrices, let us see what happens when we
include the part of the universe outside the system}.

In order to see clearly what Feynman had in mind, we use the
above-mentioned couples oscillators.  One of the oscillators is the
world in which we are interested and the other oscillator as the rest
of the universe.  There will be no effects on the first oscillator if
the system is decoupled.  Once coupled, we need a normal coordinate
system in order separate the Hamiltonian.  Then it is straightforward
to write down the wave function of the system.  Then the mathematics
of this oscillator system is directly applicable to Lorentz-boosted
harmonic oscillator wave functions, where one variable is the
longitudinal coordinate and the other is the time variable.  The system
is uncoupled if the oscillator wave function is at rest, but the
coupling becomes stronger as the oscillator is boosted to a high-speed
Lorentz frame~\cite{knp86}.

We shall then note that for two-body system, such as the hydrogen atom,
there is a time-separation variable which is to be linearly mixed with
the longitudinal space-separation variable.  This space-separation
variable is known as the Bohr radius, but we never talk about the
time-separation variable in the present form of quantum mechanics,
because this time-separation variable belongs to Feynman's rest of the
universe.

If we pretend not to know this time-separation variable, the entropy
of the system will increase when the oscillator is boosted to a
high-speed system~\cite{kiwi90pl}.  Does this increase in entropy
correspond to decoherence?  Not necessarily.  However, in 1969, Feynman
observed the parton effect in which a rapidly moving hadron appears
as a collection of incoherent partons~\cite{fey69}.  This is the
decoherence mechanism of current interest.

\section{Two Coupled Oscillators}\label{coupled}
Two coupled harmonic oscillators serve many different purposes in
physics.  It is well known that this oscillator problem can be
formulated into a problem of a quadratic equation in two variables.
To make a long story short, let us consider a system of two identical
oscillators coupled together by a spring.  The Hamiltonian is
\begin{equation}\label{hamil2}
H = {1\over 2m}\left\{p^{2}_{1} + p^{2}_{2} \right\} +
{1\over 2}\left\{K \left(x_{1}^{2} + x^{2}_{2} \right)
+ 2C x_{1} x_{2} \right\} .
\end{equation}
We are now ready to decouple this Hamiltonian by
making the coordinate rotation:
\begin{equation}\label{normal}
y_{1} = {1 \over \sqrt{2}} \left(x_{1}  - x_{2} \right) , \qquad
y_{2} = {1 \over \sqrt{2}} \left(x_{1}  + x_{2} \right) .
\end{equation}
In terms of this new set of variables, the Hamiltonian can be written as
\begin{equation}\label{eq.6}
H = {1\over 2m} \left\{p^{2}_{1} + p^{2}_{2} \right\} +
{K\over 2}\left\{e^{2\eta} y^{2}_{1} + e^{-2\eta} y^{2}_{2} \right\} ,
\end{equation}
with
\begin{equation}\label{omega}
\exp{(\eta)} = \sqrt{(K + C)/(K - C)} .
\end{equation}
Thus $\eta$ measures the strength of the coupling.
If $y_{1}$ and $y_{2}$ are measured in units of $(mK)^{1/4} $,
the ground-state wave function of this oscillator system is
\begin{equation}\label{wfc}
\psi_{\eta}(x_{1},x_{2}) = \left({1 \over {\pi}}\right)^{1/2}
\exp{\left\{-{1\over 2}(e^{\eta} y^{2}_{1} + e^{-\eta} y^{2}_{2})
\right\} } .
\end{equation}
The wave function is separable in the $y_{1}$ and $y_{2}$ variables.
However, for the variables $x_{1}$ and $x_{2}$, the story is quite
different.

The key question is how quantum mechanical calculations in the world
of the observed variable are affected when we average over the other
variable.  The $x_{2}$ space in this case corresponds to Feynman's
rest of the universe, if we only consider quantum mechanics in the
$x_{1}$ space.  As was discussed in the literature for several
different purposes~\cite{knp91,knp86}, the wave function of
Eq.(\ref{wfc}) can be expanded as
\begin{equation}\label{expan}
\psi_{\eta }(x_{1},x_{2}) = {1 \over \cosh\eta}\sum^{}_{k}
\left(\tanh{\eta \over 2}\right)^{k} \phi_{k}(x_{1}) \phi_{k}(x_{2}) .
\end{equation}
This expansion corresponds to the two-photon coherent states in Yuen's
paper~\cite{yuen76}, and the wave function of Eq.(\ref{wfc}) is
a squeezed wave function~\cite{knp91}.

\begin{figure}
\centerline{\psfig{figure=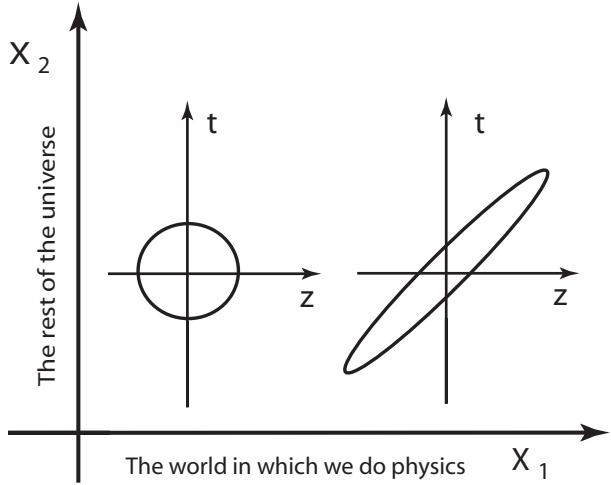,angle=0,width=80mm}}
\vspace{5mm}
\caption{Feynman's rest of the universe.  The vertical axis is not
measurable, and thus belongs to the rest of the universe.  The horizontal
axis corresponds to the world in which we do physics.}\label{frest}
\end{figure}

The question then is what lessons we can learn from the situation in
which we average over the $x_{2}$ variable.
In order to study this problem, we use the density matrix.  From this
wave function, we can construct the pure-state density matrix
\begin{equation}
\rho(x_{1},x_{2};x_{1}',x_{2}')
= \psi_{\eta}(x_{1},x_{2})\psi_{\eta}(x_{1}',x_{2}') ,
\end{equation}
If we are not able to make observations on the $x_{2}$, we should
take the trace of the $\rho$ matrix with respect to the $x_{2}$
variable.  Then the resulting density matrix is
\begin{equation}\label{integ}
\rho(x, x') = \int \psi_{\eta}(x,x_{2})
\left\{\psi_{\eta}(x',x_{2})\right\}^{*} dx_{2} .
\end{equation}
We have simplicity replaced $x_{1}$ and $x'_{1}$ by $x$ and $x'$
respectively.
If we perform the integral of Eq.(\ref{integ}), the result is
\begin{equation}\label{dmat}
\rho(x,x') = \left({1 \over \cosh(\eta/2)}\right)^{2}
\sum^{}_{k} \left(\tanh{\eta \over 2}\right)^{2k}
\phi_{k}(x)\phi^{*}_{k}(x') ,
\end{equation}
which leads to $Tr(\rho) = 1$.  It is also straightforward to compute
the integral for $Tr(\rho^{2})$.  The calculation leads to
\begin{equation}
Tr\left(\rho^{2} \right)
= \left({1 \over \cosh(\eta/2)}\right)^{4}
\sum^{}_{k} \left(\tanh{\eta \over 2}\right)^{4k} .
\end{equation}
The sum of this series is $(1/\cosh\eta)$, which is smaller than one
if the parameter $\eta$ does not vanish.

This is of course due to the fact that we are averaging over the $x_{2}$
variable which we do not measure.  The standard way to measure this
ignorance is to calculate the entropy defined as
\begin{equation}
S = - Tr\left(\rho \ln(\rho) \right) ,
\end{equation}
where $S$ is measured in units of Boltzmann's constant.  If we use the
density matrix given in Eq.(\ref{dmat}), the entropy becomes
\begin{equation}
S = \cosh^{2}\left({\eta \over 2}\right)
\ln\left(\cosh{\eta \over 2}\right)^{2} -
\sinh^{2}\left({\eta \over 2}\right)
\ln\left(\sinh{\eta \over 2} \right)^{2} .
\end{equation}
This expression can be translated into a more familiar form if
we use the notation
\begin{equation}
\tanh{\eta \over 2} = \exp\left(-{\hbar\omega \over kT}\right) ,
\end{equation}
where $\omega$ is given in Eq.(\ref{omega})~\cite{hkn89pl}.

It is known in the literature that this rise in entropy and temperature
causes the Wigner function to spread wide in phase space causing an
increase of uncertainty~\cite{hkn99ajp}.  Certainly, we cannot reach a
classical limit by increasing the uncertainty.  On the other hand, we
are accustomed to think this entropy increase has something to do with
decoherence, and we are also accustomed to think the lack of coherence
has something to do with a classical limit.  Are they compatible?  We
thus need a new vision in order to define precisely the word
``decoherence.''

\section{Time-separation Variable in Feynman's Rest of the
Universe}\label{restof}
Quantum field theory has been quite successful in terms of
perturbation techniques in quantum electrodynamics.  However, this
formalism is basically based
on the S matrix for scattering problems and useful only for physically
processes where a set of free particles becomes another set of free
particles after interaction.  Quantum field theory does not address
the question of localized probability distributions and their
covariance under Lorentz transformations.
The Schr\"odinger quantum mechanics of the hydrogen atom deals with
localized probability distribution.  Indeed, the localization condition
leads to the discrete energy spectrum.  Here, the uncertainty relation
is stated in terms of the spatial separation between the proton and
the electron.  If we believe in Lorentz covariance, there must also
be a time separation between the two constituent particles.

Before 1964~\cite{gell64}, the hydrogen atom was used for
illustrating bound states.  These days, we use hadrons which are
bound states of quarks.  Let us use the simplest hadron consisting of
two quarks bound together with an attractive force, and consider their
space-time positions $x_{a}$ and $x_{b}$, and use the variables
\begin{equation}
X = (x_{a} + x_{b})/2 , \qquad x = (x_{a} - x_{b})/2\sqrt{2} .
\end{equation}
The four-vector $X$ specifies where the hadron is located in space and
time, while the variable $x$ measures the space-time separation
between the quarks.  According to Einstein, this space-time separation
contains a time-like component which actively participates as can be
seen from
\begin{equation}\label{boostm}
\pmatrix{z' \cr t'} = \pmatrix{\cosh \eta & \sinh \eta \cr
\sinh \eta & \cosh \eta } \pmatrix{z \cr t} ,
\end{equation}
when the hadron is boosted along the $z$ direction.
In terms of the light-cone variables defined as~\cite{dir49}
\begin{equation}
u = (z + t)/\sqrt{2} , \qquad v = (z - t)/\sqrt{2} .
\end{equation}
The boost transformation of Eq.(\ref{boostm}) takes the form
\begin{equation}\label{lorensq}
u' = e^{\eta } u , \qquad v' = e^{-\eta } v .
\end{equation}
The $u$ variable becomes expanded while the $v$ variable becomes
contracted.

\begin{figure}
\centerline{\psfig{figure=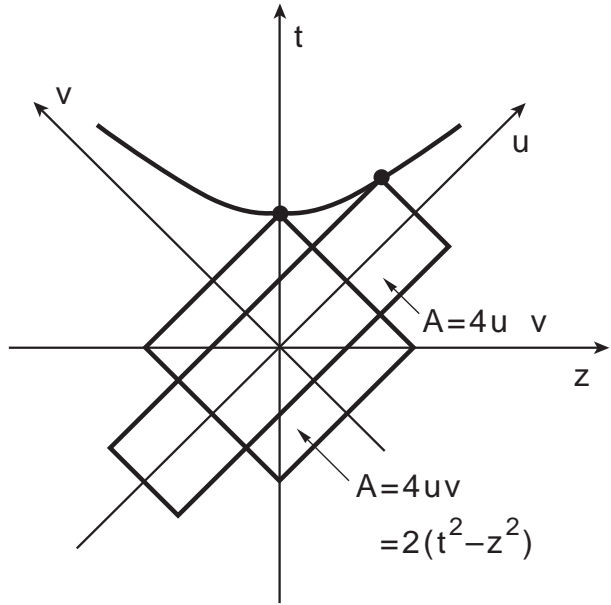,angle=0,width=80mm}}
\vspace{5mm}
\caption{Lorentz boost in the light-cone coordinate system.}\label{licone}
\end{figure}

Does this time-separation variable exist when the hadron is at rest?
Yes, according to Einstein.  In the present form of quantum mechanics,
we pretend not to know anything about this variable.  Indeed, this
variable belongs to Feynman's rest of the universe.  In this report,
we shall see the role of this time-separation variable in decoherence
mechanism.

\begin{figure}
\centerline{\psfig{figure=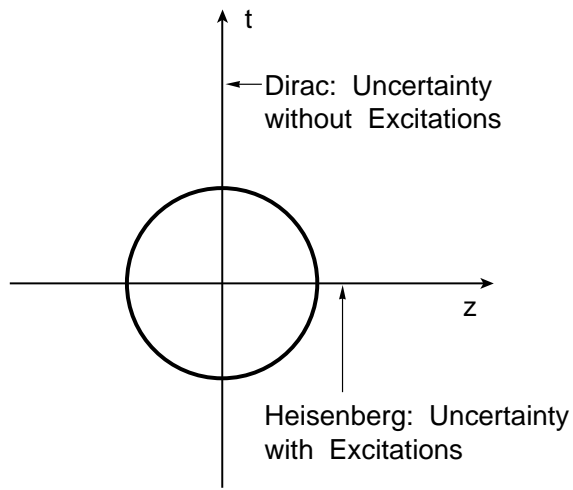,angle=0,width=80mm}}
\vspace{5mm}
\caption{Space-time picture of quantum mechanics.  There
are quantum excitations along the space-like longitudinal direction, but
there are no excitations along the time-like direction.  The time-energy
relation is a c-number uncertainty relation.}\label{quantum}
\end{figure}

Also in the present form of quantum mechanics, there is an uncertainty
relation between the time and energy variables.  However, there are
no known time-like excitations.  Unlike Heisenberg's
uncertainty relation applicable to position and momentum, the time and
energy separation variables are c-numbers, and we are not allowed to
write down the commutation relation between them.  Indeed, the
time-energy uncertainty relation is a c-number uncertainty
relation~\cite{dir27}, as is illustrated in Fig.~\ref{quantum}

How does this space-time asymmetry fit into the world of
covariance~\cite{kn73}.  This question was
studied in depth by the present authors.  The
answer is that Wigner's $O(3)$-like little group is not a
Lorentz-invariant symmetry, but is a covariant symmetry~\cite{wig39}.
It has been shown that the time-energy uncertainty applicable to the
time-separation variable fits perfectly into the $O(3)$-like symmetry
of massive relativistic particles~\cite{knp86}.

The c-number time-energy uncertainty relation allows us to write down
a time distribution function without excitations~\cite{knp86}.
If we use Gaussian forms for both space and time distributions, we
can start with the expression
\begin{equation}\label{ground}
\left({1 \over \pi} \right)^{1/2}
\exp{\left\{-{1 \over 2}\left(z^{2} + t^{2}\right)\right\}}
\end{equation}
for the ground-state wave function.  What do Feynman {\it et al.}
say about this oscillator wave function?

\begin{figure}
\centerline{\psfig{figure=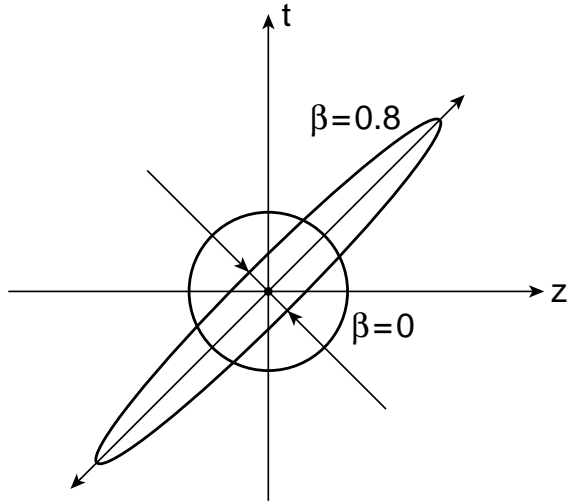,angle=0,width=80mm}}
\caption{Effect of the Lorentz boost on the space-time
wave function.  The circular space-time distribution at the rest frame
becomes Lorentz-squeezed to become an elliptic
distribution.}\label{ellipse}
\end{figure}

In his classic 1971 paper~\cite{fkr71}, Feynman {\it et al.} start
with the following Lorentz-invariant differential equation.
\begin{equation}\label{osceq}
{1\over 2} \left\{x^{2}_{\mu} -
{\partial^{2} \over \partial x_{\mu }^{2}}
\right\} \psi(x) = \lambda \psi(x) .
\end{equation}
This partial differential equation has many different solutions
depending on the choice of separable variables and boundary conditions.
Feynman {\it et al.} insist on Lorentz-invariant solutions which are
not normalizable.  On the other hand, if we insist on normalization,
the ground-state wave function takes the form of Eq.(\ref{ground}).
It is then possible to construct a representation of the
Poincar\'e group from the solutions of the above differential
equation~\cite{knp86}.  If the system is boosted, the wave function
becomes
\begin{equation}\label{eta}
\psi_{\eta }(z,t) = \left({1 \over \pi }\right)^{1/2}
\exp\left\{-{1\over 2}\left(e^{-2\eta }u^{2} +
e^{2\eta}v^{2}\right)\right\} .
\end{equation}
This wave function becomes Eq.(\ref{ground}) if $\eta$ becomes zero.
The transition from Eq.(\ref{ground}) to Eq.(\ref{eta}) is a
squeeze transformation.  The wave function of Eq.(\ref{ground}) is
distributed within a circular region in the $u v$ plane, and thus
in the $z t$ plane.  On the other hand, the wave function of
Eq.(\ref{eta}) is distributed in an elliptic region with the light-cone
axes as the major and minor axes respectively.  If $\eta$ becomes very
large, the wave function becomes concentrated along one of the
light-cone axes.  Indeed, the form given in Eq.(\ref{eta}) is a
Lorentz-squeezed wave  function.  This squeeze mechanism is
illustrated in Fig.~\ref{ellipse}.

It is interesting to note that the Lorentz-invariant differential
equation of Eq.(\ref{osceq}) contains the time-separation variable
which belongs to Feynman's rest of the universe.  Furthermore, the
wave function of Eq.(\ref{ground}) is identical to that of
Eq.(\ref{wfc}) for the coupled oscillator system, if the variables
$z$ and $t$ are replaced $x_{1}$ and $x_{2}$ respectively.
Thus the entropy increase due to the unobservable $x_{2}$ variable is
applicable to the unobserved time-separation variable $t$.


\section{Feynman's Parton Picture}\label{par}

It is a widely accepted view that hadrons are quantum bound states
of quarks having localized probability distribution.  As in all
bound-state cases, this localization condition is responsible for
the existence of discrete mass spectra.  The most convincing evidence
for this bound-state picture is the hadronic mass spectra which are
observed in high-energy laboratories~\cite{fkr71,knp86}.
However, this picture of bound states is applicable only to observers
in the Lorentz frame in which the hadron is at rest.  How would the
hadrons appear to observers in other Lorentz frames?  To answer this
question, can we use the picture of Lorentz-squeezed hadrons discussed
in Sec.~\ref{restof}.

The radius of the proton is $10^{-5}$ of that of the hydrogen atom.
Therefore, it is not unnatural to assume that the proton has a point
charge in atomic physics.  However, while carrying out experiments on
electron scattering from proton targets, Hofstadter in 1955 observed
that the proton charge is spread out~\cite{hofsta55}.
In this experiment, an electron emits a virtual photon, which
then interacts with the proton.  If the proton consists of quarks
distributed within a finite space-time region, the virtual photon will
interact with quarks which carry fractional charges.  The scattering
amplitude will depend on the way in which quarks are distributed
within the proton.  The portion of the scattering amplitude which
describes the interaction between the virtual photon and the proton
is called the form factor.

Although there have been many attempts to explain this phenomenon
within the framework of quantum field theory, it is quite natural
to expect that the wave function in the quark model will describe
the charge distribution.  In high-energy experiments, we are dealing
with the situation in which the momentum transfer in the scattering
process is large.  Indeed, the Lorentz-squeezed wave functions lead
to the correct behavior of the hadronic form factor for large
values of the momentum transfer~\cite{fuji70}.

Furthermore, in 1969, Feynman observed that a fast-moving hadron
can be regarded as a collection of many ``partons'' whose properties
do not appear to be quite different from those of the
quarks~\cite{fey69}.  For example, the number of quarks inside a
static proton is three, while the number of partons in a rapidly
moving proton appears to be infinite.  The question then is how
the proton looking like a bound state of quarks to one observer
can appear different to an observer in a different Lorentz frame?
Feynman made the following systematic observations.

\begin{itemize}

\item[a.]  The picture is valid only for hadrons moving with
  velocity close to that of light.

\item[b.]  The interaction time between the quarks becomes dilated,
   and partons behave as free independent particles.

\item[c.]  The momentum distribution of partons becomes widespread as
   the hadron moves fast.

\item[d.]  The number of partons seems to be infinite or much larger
    than that of quarks.

\end{itemize}

\noindent Because the hadron is believed to be a bound state of two
or three quarks, each of the above phenomena appears as a paradox,
particularly b) and c) together.

In order to resolve this paradox, let us write down the
momentum-energy wave function corresponding to Eq.(\ref{eta}).
If the quarks have the four-momenta $p_{a}$ and $p_{b}$, we can
construct two independent four-momentum variables~\cite{fkr71}
\begin{equation}
P = p_{a} + p_{b} , \qquad q = \sqrt{2}(p_{a} - p_{b}) ,
\end{equation}
where $P$ is the total four-momentum and is thus the hadronic
four-momentum.  $q$ measures the four-momentum separation between
the quarks.  Their light-cone variables are
\begin{equation}\label{conju}
q_{u} = (q_{0} - q_{z})/\sqrt{2} ,  \qquad
q_{v} = (q_{0} + q_{z})/\sqrt{2} .
\end{equation}
The resulting momentum-energy wave function is
\begin{equation}\label{phi}
\phi_{\eta }(q_{z},q_{0}) = \left({1 \over \pi }\right)^{1/2}
\exp\left\{-{1\over 2}\left(e^{-2\eta}q_{u}^{2} +
e^{2\eta}q_{v}^{2}\right)\right\} .
\end{equation}
Because we are using here the harmonic oscillator, the mathematical
form of the above momentum-energy wave function is identical to that
of the space-time wave function.  The Lorentz squeeze properties of
these wave functions are also the same.  This aspect of the squeeze
has been exhaustively discussed in the
literature~\cite{knp86,kn77par,kim89}.

When the hadron is at rest with $\eta = 0$, both wave functions
behave like those for the static bound state of quarks.  As $\eta$
increases, the wave functions become continuously squeezed until
they become concentrated along their respective positive
light-cone axes.  Let us look at the z-axis projection of the
space-time wave function.  Indeed, the width of the quark distribution
increases as the hadronic speed approaches that of the speed of
light.  The position of each quark appears widespread to the observer
in the laboratory frame, and the quarks appear like free particles.

\begin{figure}
\centerline{\psfig{figure=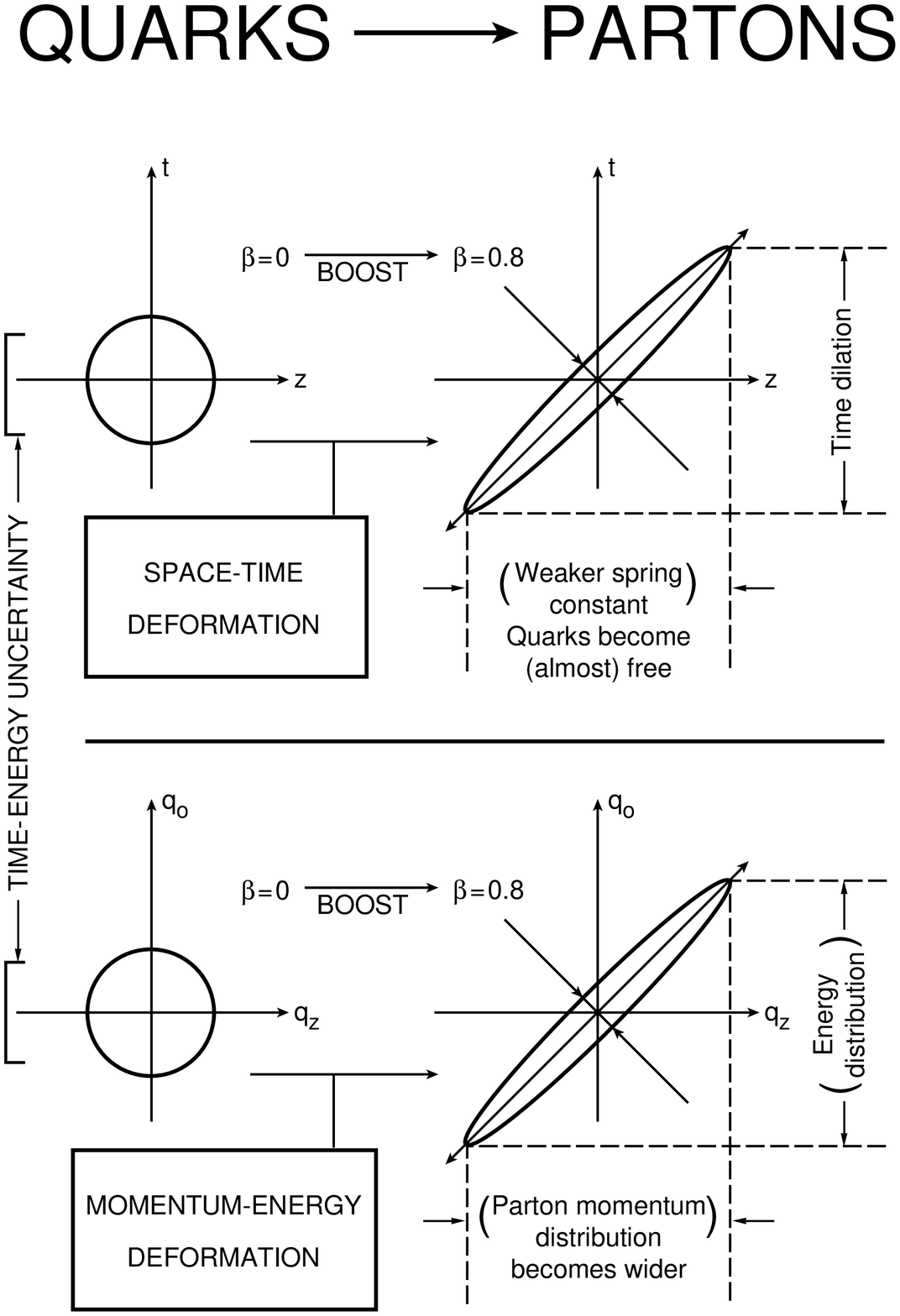,angle=0,width=70mm}}
\vspace{5mm}
\caption{Lorentz-squeezed space-time and momentum-energy wave
functions.  As the hadron's speed approaches that of light, both
wave functions become concentrated along their respective positive
light-cone axes.  These light-cone concentrations lead to Feynman's
parton picture.}\label{parton}
\end{figure}

The momentum-energy wave function is just like the space-time wave
function, as is shown in Fig.~\ref{parton}.  The longitudinal momentum
distribution becomes wide-spread as the hadronic speed approaches the
velocity of light.  This is in contradiction with our expectation from
nonrelativistic quantum mechanics that the width of the momentum
distribution is inversely proportional to that of the position wave
function.  Our expectation is that if the quarks are free, they must
have their sharply defined momenta, not a wide-spread distribution.

However, according to our Lorentz-squeezed space-time and
momentum-energy wave functions, the space-time width and the
momentum-energy width increase in the same direction as the hadron
is boosted.  This is of course an effect of Lorentz covariance.
This indeed is the key to the resolution of the quark-parton
paradox~\cite{knp86,kn77par}.

\section{Partons as Incoherent Particles}\label{incoh}

The most puzzling problem in the parton picture is that partons in
the hadron appear as incoherent particles, while quarks are coherent
when the hadron is at rest.  Does this mean that the coherence is
destroyed by the Lorentz boost?   The answer is NO, and here is the
resolution to this puzzle.

When the hadron is boosted, the hadronic matter becomes squeezed and
becomes concentrated in the elliptic region along the positive
light-cone axis, as is illustrated in Figs.~\ref{ellipse} and
\ref{parton}.  The length of the major axis becomes expanded by
$e^{\eta}$, and the minor axis is contracted by $e^{-\eta}$.

This means that the interaction time of the quarks among themselves
become dilated.  Because the wave function becomes wide-spread, the
distance between one end of the harmonic oscillator well and the
other end increases.  This effect, first noted by Feynman~\cite{fey69},
is universally observed in high-energy hadronic experiments.  The
period is oscillation increases like $e^{\eta}$.

\begin{figure}
\centerline{\epsfig{figure=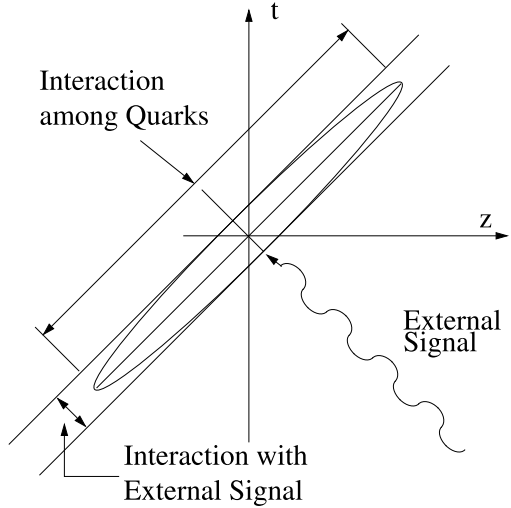,angle=0,width=80mm}}
\caption{Quarks interact among themselves and with external signal.
The interaction time of the quarks among themselves become dilated,
as the major axis of this ellipse indicates.  On the other hand, the
the external signal, since it is moving in the direction opposite to
the direction of the hadron, travels along the negative light-cone
axis.  To the external signal, if it moves with velocity of light,
the hadron appears very thin, and the quark's interaction time with
the external signal becomes very small.}\label{tdil}
\end{figure}

On the other hand, the interaction time with the external signal,
since it is moving in the direction opposite to the direction of
the hadron, travels along the negative light-cone axis, as illustrated in
Fig.~\ref{tdil}.
If the hadron contracts along the negative light-cone axis, the
interaction time decreases by $e^{-\eta}$.  The ratio of the interaction
time to the oscillator period becomes $e^{-2\eta}$.  The energy of each
proton coming out of the Fermilab accelerator is $900 GeV$.  This leads
the ratio to $10^{-6}$.  This is indeed a small number.  The external
signal is not able to sense the interaction of the quarks among
themselves inside the hadron.

Indeed, Feynman's parton picture is one concrete physical example
where the decoherence effect is observed.  As for the entropy, the
time-separation variable belongs to the rest of the universe.  Because
we are not able to observe this variable, the entropy increases
as the hadron is boosted to exhibit the parton effect.  The
decoherence is thus accompanied by an entropy increase.

Let us go back to the coupled-oscillator system.  The light-cone
variables in Eq.(\ref{eta}) correspond to the normal coordinates in
the coupled-oscillator system given in Eq.(\ref{normal}).  According
to Feynman's parton picture, the decoherence mechanism is determined
by the ratio of widths of the wave function along the two normal
coordinates.


\section{Measurement Problem}\label{measure}
In this report, we introduced a covariant model which produced the
quark model in the rest frame and the parton model where the hadron
moves with a speed close to that of light.  Unlike the case of
incoherent light arising from random phases, the present case does
not involve additional physical processes.  The decoherence we
discussed here is simply due to a frame change, and looking at the
same thing from different places.

Indeed, in the hadronic rest frame, our measurement process ignores
the existence of the time-separation variable, and our measurement
process is less than complete.  When the hadron moves very fast,
we are measuring the system through the light-cone coordinate system.
Here again, the measurement is made through only one of the two
coordinate axes. Again the measurement process is incomplete.  Thus
we have to conclude that the decoherence effect observed in Feynman's
parton picture is purely due to our observational process.  It has
nothing to with dissipation of randomness of phases.


\begin{thebibliography}{99}

\bibitem{gell64}
M. Gell-Mann, Phys. Lett. {\bf 13}, 598 (1964).

\bibitem{fey69}
R. P. Feynman,  {\it The Behavior of Hadron Collisions at Extreme
Energies}, in {\it High Energy Collisions}, Proceedings of the Third
International Conference, Stony Brook, New York, edited by C. N. Yang
{\it et al.}, Pages 237-249 (Gordon and Breach, New York, 1969).

\bibitem{fkr71}
R. P. Feynman, M. Kislinger, and F. Ravndal, Phys. Rev. D {\bf 3}, 2706
(1971).

\bibitem{fey72}
R. P. Feynman, {\it Statistical Mechanics} (Benjamin/Cummings,
Reading, MA, 1972).

\bibitem{neu32}
J. von Neumann, {\it Die mathematische Grundlagen der Quanten-mechanik}
(Springer, Berlin, 1932).  See also J. von Neumann, {\it Mathematical
Foundation of Quantum Mechanics} (Princeton University, Princeton, 1955).

\bibitem{wig39}
E. P. Wigner,  Ann. Math. {\bf 40}, 149 (1939).

\bibitem{kn73}
Y. S. Kim and M. E. Noz, Phys. Rev. D {\bf 8}, 3521 (1973).

\bibitem{knp86}
Y. S. Kim and M. E. Noz,  {\it Theory and Applications of the Poincar\'e
Group} (Reidel, Dordrecht, 1986).

\bibitem{kn77par}
Y. S. Kim and M. E. Noz, Phys. Rev. D {\bf 15}, 335 (1977).

\bibitem{kim89}
Y. S. Kim, Phys. Rev. Lett. {\bf 63}, 348 (1989).

\bibitem{kiwi90pl}
Y. S. Kim and E. P. Wigner, Phys. Lett. A {\bf 147}, 343 (1990).

\bibitem{hkn95jm}
D. Han, Y. S. Kim, and M. E. Noz, J. Math. Phys. {\bf 36},
3940 (1995).

\bibitem{yuen76}
H. P. Yuen, Phys. Rev. A {\bf 13}, 2226 (1976).

\bibitem{knp91}
Y. S. Kim and M. E. Noz, {\it Phase Space Picture of
Quantum Mechanics} (World Scientific, Singapore, 1991).

\bibitem{feyaip} This quotation is from
the Feynman page of the Emilio Segre Visual Archives
 of the American Institute of Physics.  Visit
 www.aip.org/history/esva/exhibits/feynman.htm.

\bibitem{hkn99ajp}
D. Han, Y. S. Kim, and M. E. Noz, Am. J. Phys. {\bf 67}, 61 (1999).

\bibitem{hkn89pl}
D. Han, Y. S. Kim, and Marilyn E. Noz, Phys. Lett. A {\bf 144},
111 (1989).

\bibitem{dir49}
P. A. M. Dirac, Rev. Mod. Phys. {\bf 21}, 392 (1949).

\bibitem{dir27}
P. A. M. Dirac, Proc. Roy. Soc. (London) {\bf A114}, 234 and
710 (1927).

\bibitem{hofsta55}
R. Hofstadter and R. W. McAllister, Phys. Rev. {\bf 98}, 217 (1955).

\bibitem{fuji70}
K. Fujimura, T. Kobayashi, and M. Namiki, Prog. Theor. Phys. {\bf 43},
73 (1970).

\end{thebibliography}
\end{document}